\documentclass[journal,10pt]{IEEEtran}
\usepackage{cite}
\usepackage{amsmath,amssymb,amsfonts}
\usepackage{algorithmic}
\usepackage{graphicx}
\usepackage{textcomp}
\usepackage{xcolor}
\usepackage[center]{caption}
\usepackage{todonotes}
\usepackage{url}
\usepackage{hhline}

\begin{document}

\title{A Transparent Distributed Ledger-based \\Certificate Revocation Scheme for VANETs}

\author{Andrea Tesei$^{1}$, Domenico Lattuca$^{1}$, Paolo Pagano$^{2}$, Marco Luise$^{1}$, Joaquim Ferreira$^{3}$, Paulo C. Bartolomeu$^{3}$
\thanks{$^{1}$University of Pisa, Dipartimento di Ingegneria dell'Informazione, Via G. Caruso 16, 56122 Pisa, Italy -
        {\tt\footnotesize andrea.tesei@phd.unipi.it, domenico.lattuca@phd.unipi.it, marco.luise@unipi.it}}%
\thanks{$^{2}$National Inter-university Consortium for Telecommunication (CNIT), Via Moruzzi 1, 56124 Pisa, Italy - {\tt\footnotesize paolo.pagano@cnit.it}}%
\thanks{$^{3}$Institute of Telecommunications, Universidade de Aveiro, Campus Universitário de Santiago, 3810-193 Aveiro, Portugal - {\tt\footnotesize jjcf@ua.pt, bartolomeu@ua.pt}}%

}

{}

\maketitle

\begin{abstract}
Among the available communication systems, vehicular networks are emerging as one of the most promising and yet most challenging instantiations of mobile ad-hoc network technologies. The deployment of such networks in large scale requires the enforcement of stringent security mechanisms that need to abide by the technical, societal, legal, and economical requirements of Intelligent Transportation Systems. Authentication is an effective process for validating user identity in vehicular netoworks. However, it cannot guarantee the network security by itself. Available industrial standards do not consider methods to promptly revoke misbehaving vehicles. The only available protection consists on the \textit{revocation by expiry}, which tolerates the misbehaving vehicle to remain trusted in the system for a long time (e.g. 3 months with certificate pre-loading according to EU security policy). This poses a huge yet dangerous limitation to the security of the vehicular ecosystem. In this work we propose a Distributed Ledger-based Certificate Revocation Scheme for Vehicular Ad-hoc Networks (VANETs) that harnesses the advantages of the underlying Distributed Ledger Technology (DLT) to implement a privacy-aware revocation process that is fully transparent to all participating entities and meets the critical message processing times defined by EU and US standards. An experimental validation and analysis demonstrates the effectiveness and efficiency of the proposed scheme, where the DLT streamlines the revocation operation overhead and delivers an economic solution against cyber-attacks in vehicular systems.


\end{abstract}


\IEEEpeerreviewmaketitle

\section{Introduction}\label{intro}
In the recent years, the union of transportation industry and Information and Communication Technology (ICT) become real and created new generation vehicles with strong communication capabilities, with the aim to give people better and safer driving experience. Besides the industrial growth of these technologies, Vehicular \textit{Ad-hoc} Networks (VANETs) and Intelligent Transportation System (ITS) have attracted much attention also in academia, where all research and standardization efforts are focused on creating secure frameworks that enable a set of applications in the domain of road safety, traffic efficiency and driver assistance \cite{IOTA-VPKI}. 

The ITS and VANETs in general are established with two types of communications, namely Vehicular-to-Infrastructure (V2I) communication and Vehicle-to-Vehicle (V2V) communication, that are exploited by two main entities: the device installed on vehicle, called On-Board Unit (OBU), and the Road-Side Unit (RSU) which are installed on the road. Generally speaking, through the ITS protocols and standards vehicles can exchange safety messages with V2V communications, and communicate directly with RSU in V2I communications \cite{survey-vehicular-sec}.

The open and vulnerable nature of the ITS underlying communication infrastructure, needs sophisticated security mechanisms to assure a safe real-life deployment. Furthermore, the selected security mechanism needs to face the challenging ITS security requirements, showing technical, societal, legal, and economical concerns (e.g. privacy, unlinkability, anonymity) \cite{IOTA-VPKI}.

Authentication is an effective means to ensure the security of communications. Harmonization efforts (Car2Car Communication Consortium (C2C-CC)) and standardization bodies (Institute of Electrical and Electronics Engineers (IEEE) 1609.2 Work Group (WG) \cite{ieee-16090-2019} and European Telecommunication Standards Institute (ETSI) Technical Committee (TC) ITS WG 5 with TS 102 940 \cite{etsi-ts-102-940}) have reached a consensus on the use of Vehicular Public Key Infrastructure (VPKI) to manage credentials of vehicles. With a number of Trusted Authorities (TA), VPKI systems issue different certificates to vehicles, namely, Enrolment Credentials (ECs) that are long-term certificates used to enroll vehicles in the system, and Authorization Tickets (ATs) that are pseudonyms that enable vehicle to access ITS specific applications and assure driver anonymity in the system. 
However, to assure the correct recognition of misbehaving and malicious drivers, the privacy of the driver is said to be \textit{conditional} because the union of these two types of certificate can reveal the driver identity just in case he is liable of violation \cite{SECMACE}. In fact, there exists cases in which vehicles have obtained valid credentials from VPKI but become compromised and can attack other vehicles sending false messages: in this situation, it is necessary to take advantage of new type of techniques based on attacker behaviour and message information analysis which enable \textit{misbehaviour detection} capabilities \cite{survey-revocation}. Furthermore, once a violation is recognized, a method to promptly revoke valid certificates to compromised vehicle is needed, as well as a mechanism to distribute revocation information and assure high security level of the whole system. 
Unfortunately, \textit{misbehavior detection} and \textit{revocation mechanisms} are not well covered by the latest US and EU industrial standard versions \cite{ieee-16092-2016}, \cite{ieee-16092-2017}, \cite{ieee-16092-2019}, \cite{etsi-ts-102-940}, \cite{etsi-ts-103-097}, thus limiting the deployment of Intelligent Transportation Systems that are able to guarantee the security and safety of the drivers. 

\subsection{Contribution of this paper}
We propose a novel certificate revocation mechanism encompassing the distribution of revocation information using Distributed Ledger Technology (DLT). The DLT is used to store revocation information, thus enabling vehicles to retrieve such information on-the-fly when a new secured message is received. The proposed scheme is fully compatible with the current US and EU standards and its capabilities are demonstrated with laboratory experiments thanks to the integration in our Blockchain-based VPKI named IOTA-VPKI previously presented in \cite{IOTA-VPKI} and briefly explained in Section \ref{iota-vpki-subsection}. We also show that the delay of revocation information distribution as well as the time needed to complete the revocation checking matches the standards requirements. Furthermore, we show that the revocation checking procedure delay is independent from the status of the certificate (i.e valid or revoked), as well as from the number of revoked certificates, critical requirements for a real usage of the proposed revocation scheme. We perform validation in a pseudo real environment composed by OBU, RSU and an instance of IOTA-VPKI equipped with our new revocation method. 

The rest of the paper is organized as follows: in Section \ref{background} we present the problem statement and we briefly describe relevant background; in Section \ref{proposed-scheme} we provide a detailed description of our proposed Vehicle Certificate Revocation scheme, including our previous work IOTA-VPKI, the system architecture and threat model; in Section \ref{perf-eval} the test cases and experiment settings are presented together with the obtained results; with Section \ref{conclusions} we further discuss our findings and we conclude the paper with future works.

\section{Background}\label{background}
To protect vehicles from potential attacks due to vulnerable communications and untrusted networks, a privacy-preserving authentication scheme is required. From the one hand, authentication avoid a malicious user to impersonate any authorized vehicle and broadcast false beacon and awareness messages. On the other hand, the privacy-preserving characteristic of such authentication scheme guarantees the drivers to be protected against traceability. 

Generally speaking, all methods aiming to protect vehicular systems can be broadly categorized as \textbf{proactive} or \textbf{reactive} \cite{misbehavior-survey}. Proactive security consists on the prevention of potential attackers to access the system (i.e. every mechanism that enforces a security policy, for example a VPKI), whereas reactive security consists of a \textit{detection} and \textit{reaction} steps that aims to identify and correct malicious activities within the system (i.e. misbehavior detection and revocation mechanisms). However, only the fusion of these two types of protection can assure the high level of security that is required for an ITS field deployment. 

For example, if we consider the proactive security method provided by a VPKI, the possession of valid credentials is not enough to assure that each actor of the system behave as stated in the standards and protocols. There are many cases when an issued certificate should be not valid anymore: for example, the related cryptographic material could become compromised; or changing specific fields within the issued certificates is necessary for administrative reasons; or even worst when a vehicle become compromised and starts violating registration terms or obligations. In these conditions, a reactive security approach needs to be available in order to identify misbehaving vehicles. In addition, a method to revoke the certificates issued to the compromised vehicle should be enforced. Otherwise, false messages can be potentially broadcasted to others and lead to severe accidents \cite{survey-revocation}. To avoid this situation, it is particular important that the security measures consider a fusion of proactive and reactive methods.

According to \cite{annex-2-cits-platform}, a revocation scheme must identify the following three aspects:
\begin{itemize}
    \item What set of permissions are to be revoked;
    \item What conditions must be met for the revocation of trust to occur;  
    \item What mechanism is used to communicate information about the revocation event to other vehicles enrolled and authorized in the system.
\end{itemize}

The first and last aspects are part of the so called \textit{revocation} process that is described in \cite{survey-revocation} by Wang \textit{et al.} as further divided in three stages: \textit{revocation information resolution}, \textit{revocation information distribution}, and \textit{revocation information use}. First, in order to succeed in revoking a compromised vehicle permanently, it is necessary to resolve its real identity. This stage is the \textit{revocation information resolution} and it can be done by analyzing the malicious messages sent by the compromised vehicles and its corresponding anonymous certificates. Once the real identity of the compromised vehicle has been resolved, the updated information about revoked certificates should be distributed to vehicles in the network as soon as possible, so that the vulnerability window can be minimized. This second stage is the \textit{revocation information distribution}. Finally, the revocation information is used by vehicles to determine whether or not the sender of a message should be trusted. This final stage is the \textit{revocation information use} and consists on the revocation checking process done every time a new secured message is received. This process should be as efficient as possible in order to minimize message processing latency.

Revocation schemes mainly depend on the underlying authentication available in the vehicular security architecture. As stated in Section \ref{intro}, standardization bodies have reached a consensus on the use of VPKI to authenticate and authorize vehicles in the system: this is the reason why our analysis is concentrated on revocation mechanisms that are compatible with this kind of authentication scheme. We will analyze other revocation schemes in Section \ref{related-work}.  

In general, each revocation mechanism for vehicular networks should be take in consideration the following requirements:
\begin{itemize}
    \item The distribution of the revocation information to non-revoked vehicles should happen as soon as possible, i.e. minimizing the vulnerability window \cite{optimized-crl};
    \item The revocation process should be transparent, i.e. each entity should be able to check when a specific certificate was revoked;  
    \item The revocation checking process should be as efficient as possible so that message processing latency constraints can be matched (at most 25 ms, i.e. \(\frac{1}{4}\) of the critical latency time defined in \cite{tech-report-etsi-bsa}).
\end{itemize}

\begin{figure}[htbp]
\centerline{\includegraphics[scale=1]{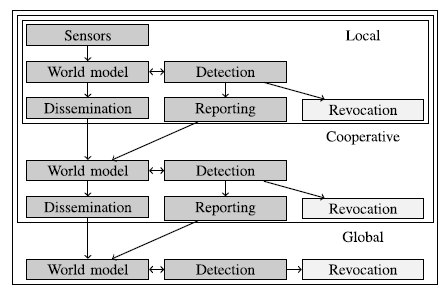}}
\caption{\centering{Different scopes at which detection mechanisms can operate \cite{survey-revocation}.}}
\label{figure-classification-misbehavior-scope}
\end{figure}

The second general aspect of a robust revocation scheme defined in \cite{annex-2-cits-platform} is related to the so called \textit{misbehavior detection}.  This mechanism is positioned in the category of reactive security mechanisms and belongs to detection step of such protection methods. According to \cite{survey-revocation}, misbehavior detection consists on the classification of the messages as correct or incorrect, where the correctness is represented by messages that reflect the real world. Misbehaving nodes are thus any node that transmit erroneous data that it should not transmit when the hardware and software are behaving as expected. Other definitions available in literature \cite{sec-requirements-vanets}, \cite{secure-vehicular-comm-systems}, \cite{illusion-attack}, \cite{security-in-vanets} often distinguishes between faulty node and malicious node: the former are those end entities that produce incorrect or inaccurate data without malicious intent (e.g. faulty software or damaged hardware); the latter are actually attacker nodes that are transmitting erroneous messages with the goal to exploit system vulnerabilities to degrade the integrity of the data exchanged in the network. A classification of detection mechanisms based on different scopes is depicted in Figure \ref{figure-classification-misbehavior-scope}, namely: \textbf{local} detection checks internal vehicle consistency; \textbf{cooperative} detection relies on collaboration between vehicles and RSUs; finally, \textbf{global} detection refers to detection that occurs in collaboration with the trust authorities. 

In all cases, once detection is done a \textit{reaction} step is executed consisting on the effective \textit{revocation} of valid credentials issued to the misbehaving vehicle. This strict dependency between revocation and misbehavior detection processes highlights that they must be taken into account for a real-life and safe deployment of ITS and VANETs in general. 

To the best of our knowledge, the only standardization activity that is running for misbehavior detection support is an european pre-standardization study available in \cite{etsi-tr-103-460}. This is a technical report that describes different approaches that can be compatible with current standards. In this report detection and reporting methods and techniques are described and different use cases are presented to describe different scenarios.
\begin{figure}[htbp]
\centerline{\includegraphics[scale=0.3]{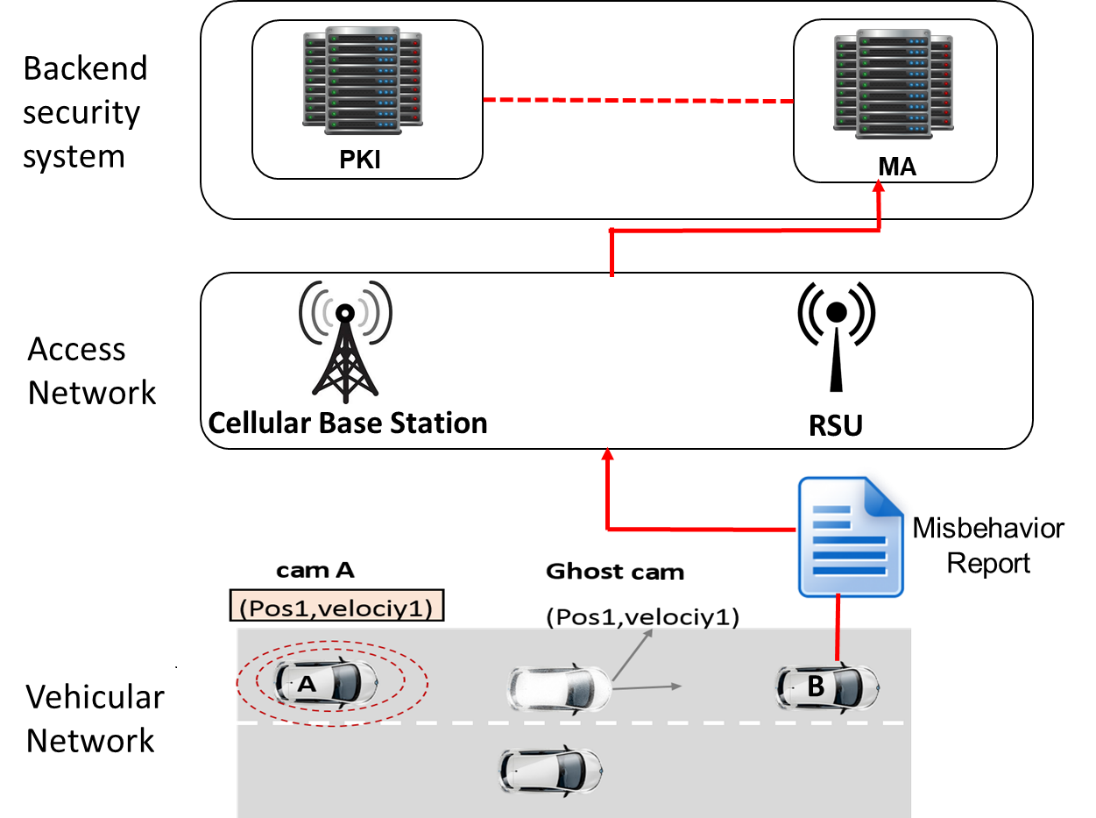}}
\caption{\centering{ETSI misbehavior reporting use case \cite{etsi-tr-103-460}.}}
\label{misbehavior-reporting-process-etsi}
\end{figure}

Figure \ref{misbehavior-reporting-process-etsi} depicts the misbehavior reporting use case proposed by ETSI in \cite{etsi-tr-103-460}, in which every ITS station can perform misbehavior detection of neighbouring vehicles and report it to the Misbehavior Authority (MA). MA is then fully integrated and connected with backend VPKI in order to trigger revocation process. Furthermore, a first draft version of the misbehaviour report message format is described in this technical report that is compatible with the existing standard on Certificate format (ETSI 103 097 \cite{etsi-ts-103-097}). 

That being said, as stated in Section \ref{intro}, besides the consensus reached by U.S. and European standards on the authentication scheme to be used in ITS, the only revocation mechanism defined is related to TAs by means of Certificate Revocation List (CRL). There is no specific revocation method for vehicles that minimize the so called \textit{vulnerability window} which represents the period in which a recognized misbehaving vehicle remains trusted by other end entities and potentially expose other drivers to damage. From US standards point of view, IEEE 1609.0 \cite{ieee-16090-2019} and IEEE 1609.2 \cite{ieee-16092-2016} (and subsequent revisions \cite{ieee-16092-2017}, \cite{ieee-16092-2019}) explicitly state that a method to distribute revocation information is still an open point that needs to be addressed also considering driver privacy concerns. From the European perspective, neither ETSI standards \cite{etsi-ts-102-940}, \cite{etsi-ts-102-941}, \cite{etsi-ts-103-097}, nor European Commission (EC) certificate policy \cite{cert-policy-eu} and norm \cite{delegated-act} for Cooperative ITS (C-ITS) consider a revocation method for vehicle certificates (i.e. ATs, and ECs): they explicitly state that no revocation method for authorization tickets and enrolment credentials is available \cite{etsi-ts-102-941}. In fact, a \textit{passive} revocation mechanism is currently used: it does not inform end entities that a specific node is to be considered revoked, but informs all relevant TAs so that when the valid credentials expire, the revoked node cannot acquire a new valid certificate. This is known as \textit{revocation by expiry} and considering that the current standards are completed by security policy \cite{cert-policy-eu} that allows a pre-loading period of three months (i.e. vehicles can request for multiple certificates before starting using them), the vulnerability window for the whole system is too high to be ready for a real life deployment.  

Thus, the currents standards only define a revocation method suitable for TA revocation that is based on Certificate Revocation Lists (CRL). CRLs are a blacklist-based revocation mechanism that is published by the TA and contains the certificates that have been revoked after a misbehavior detection process. This approach can be extended to vehicles as well but it comes with several issues. First of all, vehicular networks are delay sensitive. Hence, revocation information should be delivered to vehicles as timely as possible in order to minimize the \textit{vulnerability window} explained above. In order to match these requirements, CRLs should be distributed frequently causing high bandwidth consumption. This problem can be alleviated by using compression techniques (e.g. Bloom Filter, Delta-CRL), or by doing selective broadcasting based on geographic regions. However, this will move the issue to the RSU/OBU side thus increasing their complexity. Considering the RSU case, a geographic-aware lookup on the CRL needs to be executed to broadcast the revocation list that applies to the covered region. As fpr the OBU, it needs to implement a Bloom Filter algorithm to retrieve the revocation information and use it once a new secured message is received. Secondly, it is well known that TA are necessary for VANETs because they are responsible for vehicle registration, network maintenance, and dispute arbitration \cite{blockchain-auth-scheme}. However, their operations should be transparent to all entities participating in the network. CRLs are published by the TA without any mechanism useful to check whether a specific certificate was revoked. If a TA becomes compromised, nobody can assure the correctness of the CRLs' content. Finally, as the CRLs start growing with a huge number of revoked certificates, the revocation checking delay will increase accordingly and this may imperil the safety latency constraints defined in the standards. 

That being said, the lack of a revocation method and misbehaviour detection definition in the standards and the known issues of CRL-based approach are the main limitations for a real life and large-scale deployment of ITS technologies. That is the reason why our analysis if focused on extending the previous IOTA-VPKI work presented in \cite{IOTA-VPKI} to support a new authority named \textit{Misbehaviour Authority} (MA), which triggers the proposed vehicle certificate revocation mechanism (described in Section \ref{proposed-scheme}) and matches the requirements of ITS systems.

\section{Related Work}\label{related-work}

As stated in Section \ref{background}, CRL-based revocation schemes are the most widely used certificate revocation schemes and are proposed in many previous contributions. In \cite{survey-revocation} Wang \textit{et al.} offers a systematic investigation and wide perspective about the available revocation methods, concentrating their analysis on CRL-based approaches. They classify approaches based on the location where the revocation information has been placed: they considered revocation mechanisms implemented at the RSU side and at the Vehicle side, explaining issues and limitations of different approaches. 

Another emerging technique is based on \textit{Merkle Hash Trees} (MHTs). In \cite{ali-bc-certificateless-approach} Ali \textit{et al.} proposed a certificateless public key signature scheme that takes advantage of MHT to implement a transparent revocation mechanism exploiting Proof of Presence (PoP) and Proof of Absence (PoA) to verify the revocation status before verifying the signature on secured messages. 

An additional MHT-based approach is proposed by Ganan \textit{et al.} in \cite{ganan-et-al-mht-crl}. Here, a CRL is generated by the CA exploiting a MHT. In this case, the root hash value of such MHT is distributed to all vehicles to perform revocation checking efficiently. 

In \cite{patricia-tree-merkle-revocation} Li \textit{et al.} proposed a novel data structure named the Merkle Patricia tree (MPT) that extends the conventional blockchain structure to provide a distributed authentication scheme without the revocation list. 

Other approaches used in anonymous authentication are Group Signature (GS) as proposed in \cite{boneh-2004}, \cite{boneh-2004-2} and \cite{AAAS}. A GS-based approach implements revocation by changing the shared key used to digitally sign the secured message exchanged by vehicles. Such scheme needs to implement an efficient key management mechanism to distribute the shared key among trusted parties. 

In \cite{decentralized-key-mgmt-mechanism} Ma \textit{et al.} proposed a decentralized key management mechanism (DB-KMM) that implements automatic registration, update and revocation of user's public key based on the smart
contract technique. 

Wang \textit{et al.} proposed in \cite{identity-based-authentication-scheme} a tamper-proof device (TPD) based and privacy-preserving authentication scheme that uses a RSU to generate up-to-date secrets needed by the vehicle to generate private keys. Whenever a vehicle is recognized to be misbehaving, the RSU stops to generate secrets for that vehicle. 

Another interesting approach was proposed by Verheul \textit{et al.} in \cite{IFAL} and is named \textit{Issue First Activate Later} (IFAL). It has the ability to pre-issue pseudonym certificates that are only usable upon receiving small activation codes.

\section{Proposed Scheme}\label{proposed-scheme}
\subsection{Blockchain-based Vehicular PKI: IOTA-VPKI}\label{iota-vpki-subsection}
\begin{figure}[htbp]
\centerline{\includegraphics[scale=0.15]{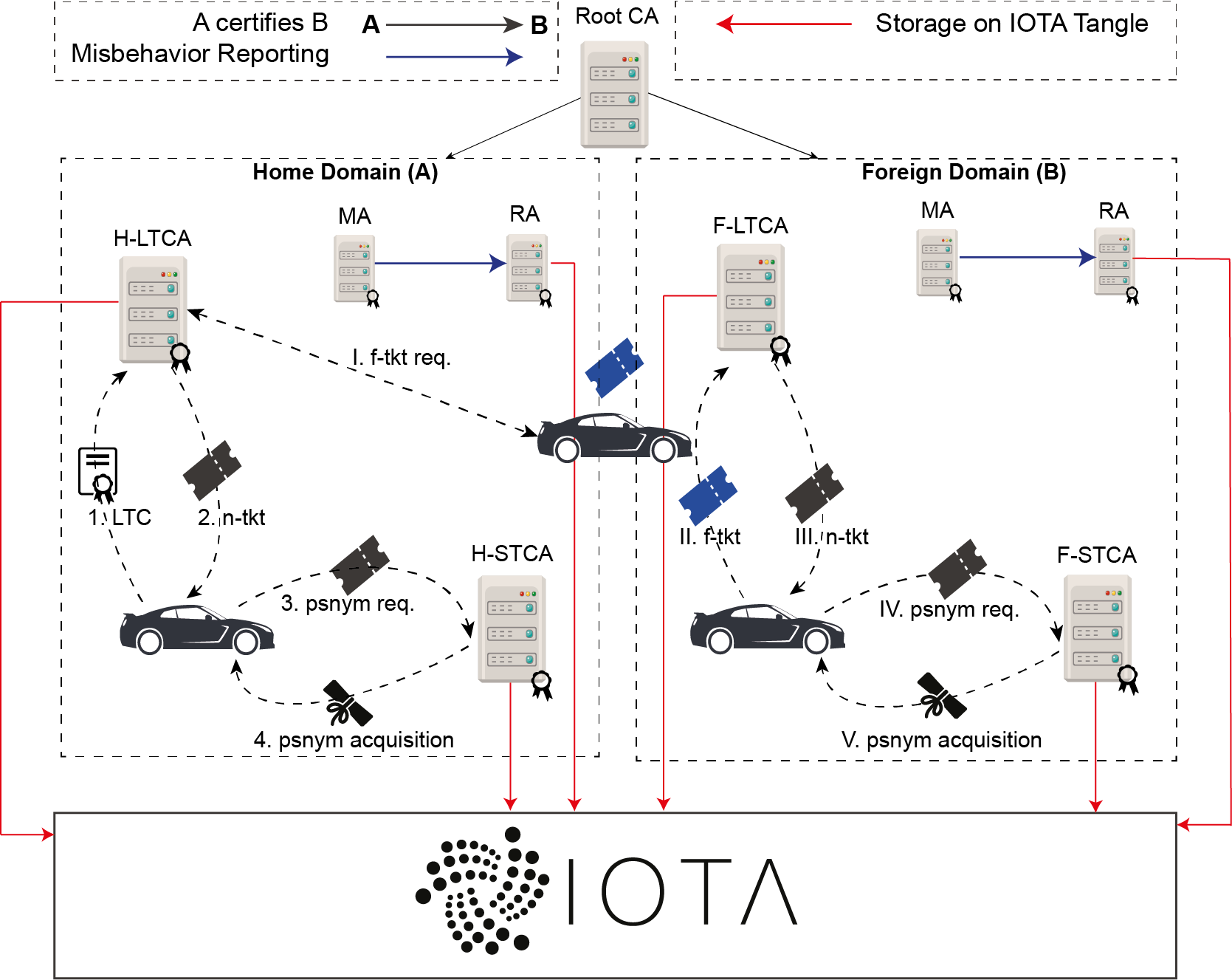}}
\caption{\centering{IOTA-VPKI Architecture: new version.}}
\label{iota-vpki-v2}
\end{figure}

We introduce IOTA-VPKI in \cite{IOTA-VPKI} as adaptation of SECMACE credential management system \cite{SECMACE}. SECMACE was proposed by \textit{Khodaei et al} and it is fully compliant with the current US and EU standards described and analyzed in Section \ref{background}. We use the SECMACE reference architecture as our starting point and we extend it with the introduction of IOTA DLT implementation as the transparent storage backend of each certificate issued to vehicles. As extensively described in \cite{IOTA-VPKI}, IOTA is a DAG-based DLT implementation well suited for IoT domain. Devices with small resource capacity can issue a transaction by communicating with the nearest neighbor IOTA Reference Implementation node (IRI).

Our VPKI architecture is composed by a set of authorities with distinct roles: the Root Certification Authority (RootCA) is the highest-level authority and represents the trust anchor of the whole system; the Long Term Certification Authority (LTCA) is responsible for vehicle enrolment in the system and Long Term Certificate (LTC - also known as Enrolment Credential) issuance; the Short Term Certification Authority (STCA) is responsible for vehicle access authorization to system applications and facilities by issuing Short Term Certificate (STC - also known as Authorization Ticket); finally the Resolution Authority (RA) can resolve a STC to a LTC identity so that a misbehaving, malfunctioning, or outdated vehicle can be promptly revoked. 

We borrow from SECMACE also the concept of \textit{home} and \textit{foreign} \textit{domains}: an \textit{Home domain} is the one where the vehicle is registered from the beginning; while a \textit{Foreign domain} is the one which a vehicle can reach after leaving its \textit{Home domain}. Namely, a \textit{domain} is defined as a set of vehicles, registered with their \textit{Home}-LTCA (H-LTCA), subject to the same administrative regulations and policies. 

When a ITS station wants to send a message, it has to acquire rights to access the C-ITS communications from the H-LTCA first by sending its own pre-registration recipe (step 1 and 2 in Figure \ref{iota-vpki-v2}). Then, it negotiates rights to access the C-ITS services from H-STCA (step 3 and 4 in Figure \ref{iota-vpki-v2}). Afterward, it digitally signs V2X messages with its private key \textit{k}$_{v}^{i}$ (corresponding to the currently valid STC), and finally sends the message if and only if all the previous steps are successfully completed (namely the vehicle is considered \textit{registered} and \textit{trusted}).

In particular, we used the IOTA feature named \textit{Masked Authenticated Message} (MAM) channels to implement data flow between TAs and vehicles. MAM channels are always managed by a \textit{channel owner} that publishes new data on such channel. In turn, devices can subscribe to the channel with read-only permissions and get data that is available \cite{iota-mam-channels}. There are three modes for MAM channels:

\begin{itemize}
\item \textit{Public}: everyone can view the data;
\item \textit{Private}: only the owner can view the data;
\item \textit{Restricted}: data is protected by a \textit{sideKey}, and the owner gives this key only to authorized viewers.
\end{itemize}

In the first IOTA-VPKI architecture version, TAs created a \textit{restricted} MAM channel to establish a certificate data flow and spread the \textit{sideKey} with pre-registered vehicles to enable them to decrypt the content of the messages (i.e., security management messages) that TAs publish on the channel. The use of MAM restricted channels act as a Group Signature (GS) based approach in which revocation can be obtained by changing the \textit{sideKey} whenever a new vehicle is revoked. 

However, whenever we start implement IOTA-VPKI we realize that MAMs have a lot of limitations. First of all, using GS approach with \textit{restricted} MAM channel is not a scalable solution because the \textit{sideKey} needs a distribution algorithm that extends the attack surface of the whole system (e.g. Man-in-the-middle attacks during key distribution phase). Secondly, the only way to perform the revocation of a vehicle consists on changing the \textit{sideKey}. This creates a potentially unbounded vulnerability window that depends on the key distribution algorithm execution time. Lastly, by using a core feature of a specific DLT implementation would lock-in our architecture to this specific technology becoming dependent on its future evolution.

For this reason we completely abandoned the MAM approach and embraced the concept of IOTA zero-value transactions \cite{iota-zero-value} as the mean to establish the data flow between CAs and vehicles. As depicted in Figure \ref{iota-vpki-v2}, we have changed MAM messages related to VPKI operations with general storage in IOTA Tangle exploiting zero-value transactions. The only known limitation of zero-value transactions is related to the so called IOTA Snapshot process: when this process takes place, all zero-value transactions get reset. Hence, the stored information gets lost. To overcome this issue, an IOTA \textit{permanent} node is required to store the IOTA-VPKI zero-value transactions and let them remain available even after a Snapshot process. This kind of permanent node is called \textit{Chronicle} that is a \textit{permanode} solution which takes transactions from a node and stores them in a distributed database that is secure and scales well \cite{iota-chronicle}. 

In the new IOTA-VPKI we have also introduced a new authority named \textit{Misbehavior Authority} (MA) that is responsible of Misbehavior Detection process that is needed to trigger the revocation process. The way that the MA performs the detection of misbehaving vehicles is out of the scope of the present paper and is currently under investigation. This new approach exploits the basic transactions of a DLT implementation, ensuring that the proposed architecture can be extended to support different available DLT technologies, thus avoiding the lock-in problem described above. Furthermore, it enables the implementation of a general purpose, transparent, and privacy-aware revocation mechanism that also supports the \textit{active revocation} of misbehaving vehicles while taking advantage of the DLT to store and distribute revocation information in a very short time. The proposed revocation mechanism is further discussed in the next subsection.

\subsection{Vehicle Certificate Revocation Mechanism}\label{revocation-mechanism}
\begin{figure*}
\includegraphics[width=\textwidth]{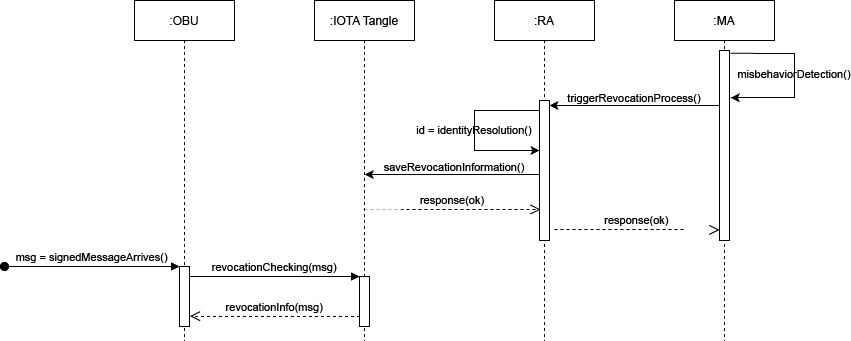}
\caption{\centering{Revocation mechanism flow diagram.}}
\label{revocation-flow-diagram}
\end{figure*}

Starting from the latest version of IOTA-VPKI architecture depicted in Figure \ref{iota-vpki-v2}, we developed a novel certificate revocation mechanism fully compliant with EU and US standards as well as European Certificate Policy \cite{cert-policy-eu}. This certificate revocation method takes advantage of the DLT to store and distribute revocation information to all participants in the network.

As described in Section \ref{background}, the revocation process starts whenever a vehicle is detected to be malfunctioning or misbehaving. Regardless of what is the intent (i.e. faulty or malicious node), the valid credentials previously issued to the misbehaving vehicle need to be revoked and all other participants need to be promptly informed. To do so, we considered the unique representation of a certificate defined in IEEE 1609.2 (clause 6.4.3) \cite{ieee-16092-2016} which defines the so called \textit{whole-certificate hash algorithm}, namely the standard way to encode the whole certificate with secure hash function. This algorithm produces an hash value of 3, 8, or 10 bytes (i.e. ASN.1 definitions: \textit{HashedId3}, \textit{HashedId8}, and \textit{HashedId10} respectively). Given the standard hash representation of a generic certificate, we analyzed the current IOTA address format defined in \cite{iota-address-format}, and used the IOTA official library to encode the certificate hash value in a valid \textit{tryte} IOTA address format. In this way, whenever a new misbehaving vehicle is detected by the MA, the RA simply calculates the \textit{tryte-encoded} representation of the given certificate hash value and attaches a new \textit{zero-value} transaction to the IOTA Tangle. The transaction also contains the  \textit{tryte-encoded} value signed by the RA: this is needed to avoid Denial of Service (DoS) attacks from malicious actors trying to revoke vehicles without authorization. 

On the OBU side, whenever a new secured message is received the Security Entity (SE) checks the revocation status of the corresponding certificate by accessing IOTA Tangle at the address obtained by \textit{tryte-encoding} the sender's certificate: if at least a transaction exists, the certificate were revoked and the message cannot be trusted. The flow diagram of the revocation process that starts from a misbehavior detection event is depicted in Figure \ref{revocation-flow-diagram}.

Since the revocation checking consists only on the \textit{tryte} value calculation of the sender's certificate hash value, the time needed to check revocation status is constant and it is independent from the number of the revoked certificates. 

The revocation checking process happens directly by accessing the IOTA Tangle without V2I communication: this avoids to increase the traffic through VPKI elements and increases the availability of revocation information that is always accessible by vehicles and trust authorities. Just in case a vehicle has a limited access to the network and cannot access the IOTA Tangle directly, it can exploit Peer-to-Peer (P2P) communication to delegate revocation checking process to a neighbouring vehicle. A similar P2P communication process is defined in IEEE 1609.2 \cite{ieee-16092-2016} as well as in ETSI 102 941 \cite{etsi-ts-102-941} for P2P Certificate Distribution (P2PCD) used when end entity has not access to cellular network connection. In the worst case scenario, when neighbouring vehicles also do not have access to the Web, the revocation checking process cannot be performed. Therefore, to avoid taking into consideration messages from vehicles that may have been compromised, the received secured message is ignored.

Apart from the RA and the MA that are responsible for misbehavior detection and the revocation process, other trust authorities are important as well in order to resolve the identity of the misbehaving vehicle starting from anonymous certificate (i.e. STC). In fact, they are involved in the revocation process because the first step that the RA needs to complete is the resolution of the compromised vehicle identity. Furthermore, once the revocation information is stored on the IOTA Tangle, they need to be promptly informed so that a new valid certificates to the revoked vehicle can not be issued. 

\begin{figure*}
\includegraphics[width=\textwidth]{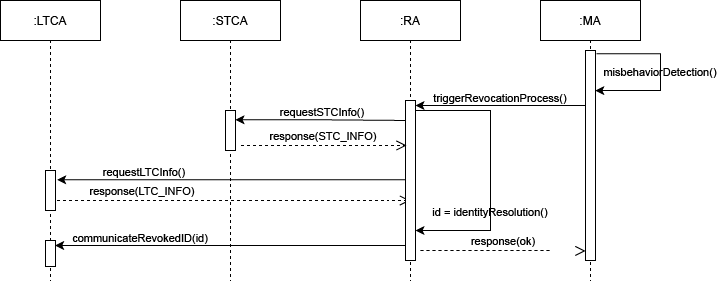}
\caption{\centering{Identity resolution of misbehaving vehicle flow diagram.}}
\label{revocation-id-resolution}
\end{figure*}

In Figure \ref{revocation-id-resolution} is depicted the flow diagram that models the identity resolution of a misbehaving vehicle. Once the MA sends a misbehavior report to RA, the STCA is contacted by the RA to obtain information about the STC used by the misbehaving vehicle to sign V2X messages. During this step, the STCA looks inside the list of issued certificates searching for the LTC that was used by the compromised vehicle to obtain the STC. With this information, the RA contacts the LTCA to retrieve information about the LTC received by the STCA. At this point, the RA can match the information received by the two trust authorities in order to extract the so called \textit{canonical identifier} of the compromised vehicle. Hence, the identity resolution process is completed and the RA needs to communicate to the LTCA that the retrieved \textit{canonical identifier} needs to be banned from the system and cannot obtain new valid certificates. This communication does not happen with the STCA because, as stated in the standards, every STC issuance process depends on the acceptance of the LTCA that is contacted by the STCA to validate the LTC provided by the vehicle wishing to obtain a new short-term anonymous certificate. 

Generally speaking, the proposed revocation scheme completely matches the requirements discussed in Section \ref{background}: it realizes the distribution of the revocation information by exploiting the DLT underlying technology; it guarantees zero-overhead on vehicle during the revocation checking process; and, thanks to the DLT underlying storage, it also provides \textit{transparency} to the whole process.

Our revocation mechanism was tested in a pseudo-real deployment set up with a IOTA Private Tangle, an instance of IOTA-VPKI extended with this revocation scheme, a RSU simulated on a desktop workstation, and an OBU that run real software on embedded system equipped with a System-on-Module (SOM). The results of such simulations are better than the ones provided by SECMACE comparative analysis \cite{SECMACE} and will be further discussed in Section \ref{perf-eval}.

\subsection{Threat Model}\label{threat-model}
To complete our analysis, we considered ETSI Threat, Vulnerability and Risk Analysis (TVRA) report in order to analyze possible threats and vulnerabilities in our proposed solution. The TVRA method is used to identify risks by isolating vulnerabilities of the system, assessing the likelihood of malicious attacks on that vulnerability and determining the impact that such an attack will have on the whole system \cite{etsi-tr-102-893}. The report identified the following security objectives:

\begin{itemize}
    \item \textit{Confidentiality}: every information sent to or from an authorized entity should not be revealed to any party not authorized to receive the information. Furthermore, it should not be possible for unauthorized party to deduce the location, identity, and route taken by a vehicle based on communication traffic analysis;
    \item \textit{Integrity}: every information sent to or from an authorized entity should be protected from unauthorized modification, deletion, malicious modification or manipulation during transmission;
    \item \textit{Availability}: access to ITS services by authorized entities should not be prevented by malicious activity within the system;
    \item \textit{Accountability}: it should be possible to audit all changes to security parameters and applications like updates, additions, and deletions;
    \item \textit{Authenticity}: it should not be possible for an unauthorized user to impersonate authorized entity when communicating with other authorized party.
\end{itemize}

For each security objective, the report lists a set of vulnerabilities that may be exploited from malicious actors to degrade the security level of the whole system. We selected threats that may apply to the proposed revocation scheme as well as to the IOTA-VPKI architecture. 

\begin{itemize}
    \item \textit{Denial of Service (DoS) attack}: this kind of attacks can substantially degrade the availability of the whole system with, for example, the intentional introduction of high volume of messages that result in a limitation of access to ITS services by authorized and harmless vehicles. The IOTA-VPKI as well as the proposed revocation scheme take advantage of the DLT technology to store VPKI operations and revocation information. In this condition, if an attacker wants to degrade the information availability the only way is to attack the underlying DLT technology directly which is challenging \cite{dag-based-security-analysis};
    \item \textit{Manipulation attack}: malicious modification or manipulation of credential management information can severely limit the integrity security objective. The IOTA-VPKI and the proposed revocation scheme provide protection against manipulation attacks by requiring cryptographic signatures within each message sent to authorized vehicles and stored in the DLT. Without the RA and TA private keys it is not feasible for an attacker to modify credentials or revocation information stored in the DLT or sent to authorized vehicles;
    \item \textit{Insertion of information attack}: insertion of crafted malicious information aims to degrade the integrity of the information available in the system. If the malicious information will be trusted by harmless vehicles to take decisions, the consequences can weaken the correctness of security operations and management, or worse, endanger the life of the passengers. Revocation information contains the RA signature so it can only be stored on the IOTA Tangle by the authorized authority (RA). Signature cryptography prevents malicious actors to impersonate the RA thus avoiding unauthorized insertion of information within the system;
    \item \textit{Man-In-The-Middle attack}: a malicious actor can degrade the integrity of the VPKI system by routing vehicle's credential requests to the attacker that, in this way, impersonates a TA. Furthermore, an attacker can eavesdrop the revocation verification requests through the IOTA Tangle and attempt to make the revoked vehicle trusted by other vehicles. The proposed revocation mechanism prevents such scenario by inserting the RA signature within the revocation information stored on the IOTA Tangle: this way any harmless vehicle can check the signature and the integrity of the revocation information.
\end{itemize}

In summary, the new version of IOTA-VPKI architecture and the revocation mechanism proposed in this paper is robust against \textit{Denial of Service}, \textit{Manipulation}, \textit{Malicious insertion of Information}, and \textit{Man-In-The-Middle} attacks while meeting the \textit{Availability}, \textit{Integrity}, \textit{Authenticity}, \textit{Accountability} and \textit{Confidentiality} security objectives and requirements stated in ETSI TVRA report \cite{etsi-tr-102-893}.

\begin{figure*}
\includegraphics[width=\textwidth]{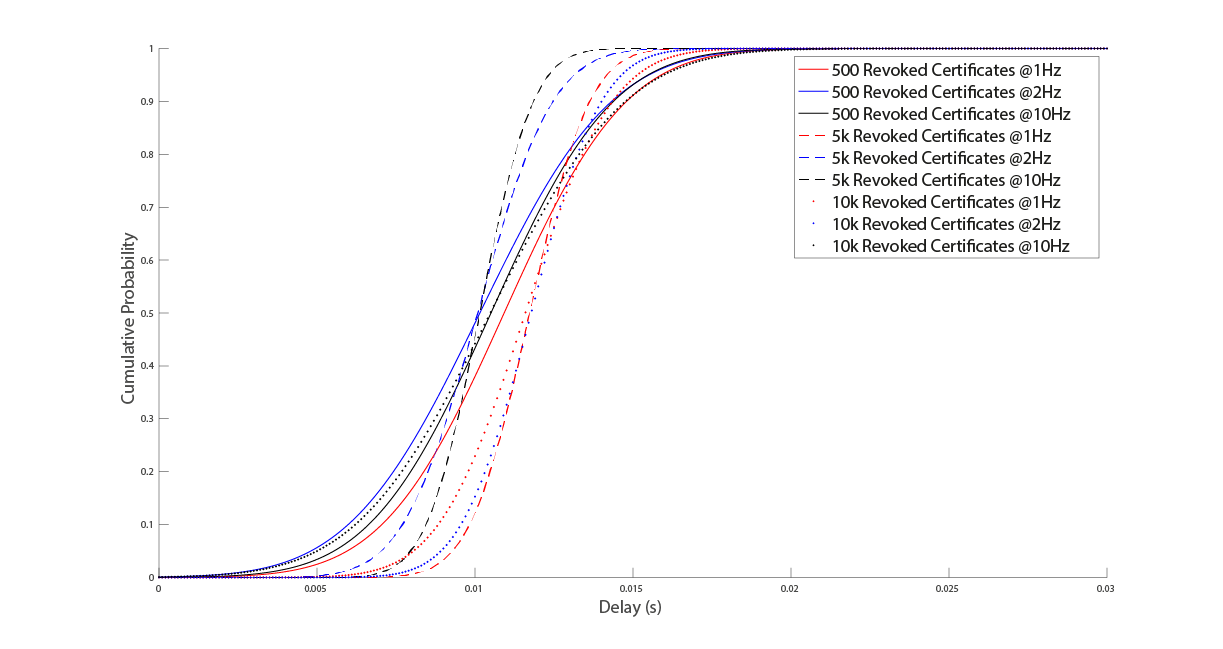}
\caption{\centering{CDF of revocation checking delay for the different test runs (OBU side).}}
\label{revocation-check-delay-cdf}
\end{figure*}

\section{Performance Evaluation}\label{perf-eval}

\subsection{Experiment Settings}
To demonstrate the effectiveness of the proposed solution, a pseudo-real environment was created in our laboratory and multiple test runs in different conditions were conducted to stress the revocation method and evaluate if it meets the requirements discussed above. 

We used two workstation with 3.0-GHz Intel Core i5 CPU and 8-GB RAM to deploy an instance of an RSU equipped with the IOTA Reference Implementation (IRI) node that acted as gateway towards the IOTA ledger for issuing transactions, and an instance of IOTA-VPKI extended with the proposed certificate revocation scheme. The two workstation were connected to each other through a 1-Gb/s switch. To simulate the OBU we use our proprietary board equipped with SOM NXP i.MX 8M Quad-core (4 x Cortex™-A53 1.5GHz) and 4GB RAM, which simulated the revocation checking procedure in different runs.

To avoid interference with public transactions, we have also deployed an instance of Private IOTA Tangle on IOTA-VPKI instance \cite{iota-private-tangle-howto}. The Private Tangle instance was executed with basic settings equal to Devnet \cite{devnet-conf}. However, the proposed method is perfectly compatible with the public IOTA Tangle instance.

\subsection{Experimental Results}

In each run we simulated that an OBU receives signed messages at different frequency from different senders. Upon receiving a new message, the OBU first checks for the status of the sender's certificate by calculating the IOTA tryte address corresponding to the hash representation of the given certificate. If there exists at least one transaction on the obtained address, the OBU ignores the received message and goes on processing a new message. The revocation checking is done by exploiting the IRI node deployed on the RSU. This process was implemented in Python for both revocation execution and revocation checking process.

In order to simulate real message frequencies, we have considered what is stated by ETSI in the Basic Set of Applications Definitions \cite{tech-report-etsi-bsa}: the different frequencies according to different message types and various conditions that may occur on the road are 1 Hz, 2 Hz, and 10 Hz. The use of multiple message frequencies was required to study the effectiveness of the proposed certificate revocation scheme under different road conditions (e.g. vehicles increase frequency in hazardous situation). Furthermore, it allowed analysing if the latency of the revocation process checking is independent on the available message frequency.

Finally, to demonstrate that the latency of revocation process checking is also independent from the status of the certificate (i.e. valid or revoked), we set up each run with half of the issued certificates in revoked status. We know that in a real situation this condition is not feasible because the number of valid certificates is generally higher than of those revoked. However, to guarantee that the OBU processes a message from a valid or revoked vehicle with equal probability that setting was mandatory. 


 Figure \ref{revocation-check-delay-cdf} illustrates the Cumulative Distribution Function (CDF) of the revocation checking procedure with different numbers of revoked certificates and distinct message frequencies. As summarized in Table \ref{table:revocation-checking-stats}, in the first run (500 revoked certificates) there is a 95\% of probability that the delays are lower or equal than 16.1 ms when considering a 1 Hz message frequency. Increasing the message frequency has almost no impact on the delay for the 95\% probability:  15.6 ms for both 2 Hz and 10 Hz message frequencies. The average delay value is slightly higher than 10 ms for all message frequencies in the first run, while the maximum delay value of 31.5 ms was measured at a 2 Hz message frequency.

In the second and third runs the measurements are very close to the first one. In fact, with 5K revoked certificates the delays are lower or equal than 14.2 ms, 13.2 ms, and 12.4 ms in 95\% of the cases regarding frequencies of 1, 2 and 10 Hz, respectively. Similarly, in the third run (10K revoked certificates) the Pr\{\textit{t} $\leq$ 15.2 ms\} = 0.95 for 1 Hz message frequency and remains close to this value with increased frequency values (i.e., 14.7 ms for 2 Hz and 16.1 ms for 10 Hz). The average values are still very close to the first run, namely around 11 ms (1 Hz), 10 ms (2 Hz and 10Hz). Finally, the worst case scenario in the second and third runs is equal to 29.8 ms and was measured at a 1Hz message frequency with 10K revoked certificates. Results confirm that the revocation checking delay is independent from the number of issued and revoked certificates as well as from the message frequency. Furthermore, the majority of the measurements (95\%) are significantly lower than 25 ms, which demonstrates that the proposed revocation method is compatible with the requirements defined by ETSI in \cite{tech-report-etsi-bsa}.

\begin{table}[ht!]
\begin{center}
    \begin{tabular}{ |c|c|c|c| }
        \hline
        \multicolumn{4}{|c|}{Frequency 1 Hz}\\
        \hline 
            \# Revoked & Average & Maximum & Pr\{t $\leq$ x\} = 0.95 \\
        \hline
            500 & 10.9 ms & 31 ms & 16.1 ms \\
        \hline
            5K & 11.7 ms & 24 ms & 14.2 ms \\
        \hline
            10K & 11.6 ms & 29.8 ms & 15.2 ms \\
        \hline \hline
        \multicolumn{4}{|c|}{Frequency 2 Hz}\\
        \hline 
            500 & 10.2 ms & 31.5 ms & 15.6 ms \\
        \hline
            5K & 10.1 ms & 22.7 ms & 13.2 ms \\
        \hline
            10K & 11.8 ms & 24.9 ms & 14.7 ms \\
        \hline \hline
        \multicolumn{4}{|c|}{Frequency 10 Hz}\\
        \hline
            500 & 10.5 ms & 25.2 ms & 15.6 ms \\
        \hline
            5K & 10.1 ms & 25 ms & 12.4 ms \\
        \hline
            10K & 10.5 ms & 26.4 ms & 16.1 ms \\
        \hline
    \end{tabular}
\end{center}
\caption{Revocation checking delay statistics.}
\label{table:revocation-checking-stats}
\vspace{-1mm}
\end{table}

\begin{figure}
\includegraphics[scale=0.47]{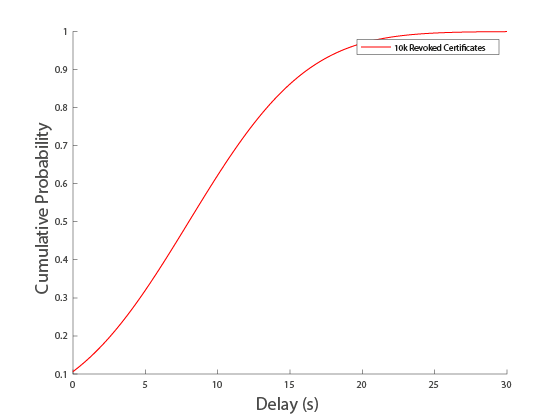}
\caption{\centering{CDF of revocation process delay (IOTA-VPKI side).}}
\label{revocation-process-delay-cdf}
\end{figure}

\begin{table}[ht!]
\begin{center}
    \begin{tabular}{ |c|c|c|c| }
        \hline 
            \# Revoked & Average & Maximum & Pr\{t $\leq$ x\} = 0.95 \\
        \hline
            10K & 8 s & 82.96 s & 18.57 s \\
        \hline 
    \end{tabular}
\end{center}
\caption{Revocation process delay statistics.}
\label{table:revocation-process-stats}
\vspace{-1mm}
\end{table}

Figure \ref{revocation-process-delay-cdf} illustrates the CDF of the revocation process delay with 10K measurements. This delay represents the time to attach a zero-transaction on IOTA Tangle at the address derived by the hash value of the certificate to be revoked. As shown in Table \ref{table:revocation-process-stats}, the \textit{vulnerability window} of the proposed scheme is equal or lower than 18.57 seconds in 95\% of the cases, a very short interval when compared to those in \textit{revocation by expiry} schemes. However, in the worst case scenario a misbehaving vehicle remains trusted in the system for 82.96 seconds, which means that at the maximum message frequency available (i.e. 10Hz), an attacker can send a maximum of 829 messages before the revocation information arrives at the other harmless participants. Nevertheless, this value is still much lower than what occurs in the \textit{vulnerability window} exposed by the current standards.   

\subsection{Comparison With Other Implementations}
\begin{table}[ht!]
\begin{center}
\begin{tabular}{ |p{2.5cm}||p{5cm}| }
 \hline
& \hfil{Delay} \\
 \hline
 VeSPA \cite{vespa}   & \hfil $\approx$ 800 ms     \\
 SEROSA \cite{serosa} &  \hfil $\approx$ 800 ms   \\
 SECMACE \cite{SECMACE} & \hfil $\leq$  75 ms \\
 IOTA-VPKI    & \hfil $\leq$ 25 ms \\
 \hline
\end{tabular}
\end{center}
\caption{Revocation checking delay comparison.}
\label{table-comparison-delay-revocation}
\vspace{-1mm}
\end{table}
We considered other implementations that provide also performance evaluation for revocation mechanism. They are all related to credential management systems which support active vehicle certificate revocation. Other proposals available in the literature \cite{puca}, \cite{preserve}, \cite{c2c-pilot}, support only revocation by expiry, so they are not directly comparable with our revocation mechanism. Table \ref{table-comparison-delay-revocation} shows the delay of revocation checking comparisons from the OBU side. The results confirm a significant performance improvement of our revocation scheme over prior art, namely a 3-fold improvement over SECMACE \cite{SECMACE}, best case out of the analyzed solution. It is worth notice that SECMACE solution take advantage of OCSP verification protocol \cite{ocsp} but the verification is done after retrieving the CRL from the LTCA: the performance evaluation available in \cite{SECMACE} states that for a CRL with 100.000 pseudonyms the latency for obtaining a CRL is less than 1500 ms. In our solution this delay is completely absent since vehicle access revocation information directly on IOTA Tangle.

Furthermore, the delay of our proposed scheme is independent from the number or revoked certificates, which is not the case for other considered solutions \cite{vespa}, \cite{serosa}. 

\section{Conclusion}\label{conclusions}
 In this paper we have presented a new revocation method for VANETs that assures the transparency of the revocation process. 
 The solution was based on our previous work IOTA-VPKI \cite{IOTA-VPKI}, a DLT-based VPKI compliant with ETSI standards. Starting on our first architecture version, we enhanced the Resolution Authority (RA) to support the proposed revocation mechanism taking advantage of the Distributed Ledger Technology (DLT) to save revocation information. Furthermore, we extended the IOTA-VPKI architecture with a \textit{Misbehavior Authority} (MA) that is responsible for the detection of malicious or misbehaving vehicles so that they are promptly revoked. Once a vehicle is recognized to be compromised by the MA, the RA is triggered to publish the revocation information on the IOTA Tangle ledger with a zero-value transaction. In parallel, the RA starts the identity resolution process to retrieve the vehicle identity and communicate it to other CAs, indicating that no more valid credentials can be issued to the compromised vehicle. Once an OBU receives a new secured message it checks the IOTA Tangle for a revocation transaction considering the anonymous certificate used to sign the secured message: if at least one zero-value transaction signed by a trusted CA exists at the address represented by the anonymous certificate, the message must be ignored.

To complement our proposal, a threat model based on ETSI TVRA report \cite{etsi-tr-102-893} is considered with a discussion of the IOTA-VPKI robustness against Denial of Service, Manipulation, Malicious insertion of Information, and Man-In-The-Middle attacks.
 
Experimental results document that 95\% of the revocation checking delays are lower or equal than 16 ms, which demonstrates the feasibility of the proposed revocation method and compatibility with the requirements defined by ETSI in \cite{tech-report-etsi-bsa}, enabling its use in realistic ITS environments. Furthermore, results confirm that the revocation checking delay is independent from the status of the certificate (i.e., issued or revoked) as well as from the total number of issued certificates. This result confirms that the proposed scheme overcomes the issues found in CRL-based approaches  available in current US and EU standards. Finally, the underlying IOTA DLT storage guarantees the transparency of the whole revocation process, as well as the permanent availability of revocation information that does not require complex distribution protocols. 

In turn, the RA revocation process delay measurements demonstrate that the \textit{vulnerability window} of the proposed revocation scheme is lower than 18.57 seconds on the majority (95\%) of the measurements. This is an important improvement with respect to the \textit{vulnerability window} available in the current standards that can reach 3 months according to the latest European security policy (certificate pre-loading case \cite{cert-policy-eu}). Even considering the worst case scenario (82.96 seconds), the \textit{vulnerability window} remains much lower than the one exposed by \textit{passive} revocation mechanism such as \textit{revocation by expiry}.

In conclusion, we consider that our solution can be proposed as an extension of the current EU and US standards in order to mitigate the risks and lower the \textit{vulnerability window} of the current ITS security architecture. The proposed revocation scheme perfectly matches the requirements of VANETs, and it is ready to be used in a real and large scale ITS deployment.

As future works, we plan to contribute to the standardization effort with ASN.1 modules describing the data structures and messages used by our revocation mechanism. Furthermore, we plan to test the IOTA-VPKI system integrated with the proposed solution in two real testbeds: the first is available at the Livorno seaport and highway (Italy); the second is available in a Smart City context at Aveiro (Portugal). With these pilot sites we will demonstrate the effectiveness of the proposed solution in an European large scale testbed embodying a realistic ITS environment.

 











\end{document}